\documentclass[aps,twocolumn,floatfix,prl,letterpaper,superscriptaddress]{revtex4-1}
\usepackage{amsmath,hyperref,amssymb,graphicx,epsf,epstopdf,dsfont,float,scrextend,color,mathtools,ulem}
\allowdisplaybreaks
\newcommand{\beq}{\begin{equation}}
\newcommand{\eeq}{\end{equation}}
\newcommand{\bpm}{\begin{pmatrix}}
\newcommand{\epm}{\end{pmatrix}}

\newcommand{\ba}{\begin{array}{l}}
\newcommand{\ea}{\end{array}}

\newcommand{\B}[1]{\boldsymbol{#1}}

\begin{document}
\title{Combing a double helix}
\date{\today}

\author{Thomas Bolton Plumb-Reyes}
\affiliation{School of Engineering and Applied Sciences,
Harvard University, Cambridge, MA 02138}
\author{Nicholas Charles}
\affiliation{School of Engineering and Applied Sciences,
Harvard University, Cambridge, MA 02138}
\author{L. Mahadevan}
\affiliation{School of Engineering and Applied Sciences, Departments of Physics and Organismic and Evolutionary Biology,
Harvard University, Cambridge, MA 02138}
\email{lmahadev@g.harvard.edu}

\begin{abstract}
Combing hair involves brushing away the topological tangles in a collective curl.   Using a combination of experiment and computation, we study this problem that naturally links topology, geometry and mechanics. Observations show that the dominant interactions in hair are those of a two-body nature, corresponding to a braided homochiral double helix. Using this minimal model, we study the detangling of an elastic double helix via a single stiff tine that moves along it, leaving two untangled filaments in its wake.  Our results quantify how the forces of detangling are correlated with the magnitude and spatial extent of the link density, a topological quantity, that propagates ahead of the tine. This in turn provides a measure of the maximum characteristic length of a single combing stroke, and thus the trade-offs between comfort, efficiency and speed of combing in the many-body problem on a head of hair. 
\end{abstract}

\maketitle

Long-haired people are familiar with a well-known strategy for combing their hair: start combing away the tangles close to the free hair ends, and work steadily upward towards the scalp. This allows for the untangling of a  collective \textit{curl}  to proceed more efficiently from the free end, minimizing pain but at the expense of time.  But how does a comb work its way through a curl? This quotidian problem which lies at the intersection of mechanics, geometry and topology  has many cousins---the carding of textiles and felts, and the spontaneous tangling and detangling of polymers in a flow, of flux lines in superconductors and of magnetic fields in solar coronae. In the context of hair, there has been a recent resurgence of interest in characterizing the effective properties of fiber assemblies and packings \cite{Hearle,Pan, Panaitescu, Kabla, Leaf, Goldstein2018, Durand2009, Sacks2017}, inspired by simple curiosity, technological applications to fields such as robotics, and even the need to model hair animation \cite{Ward2007,Bao2017}. Much of these studies neglect individual hair-hair interactions, and little is known about the dynamics of detangling in complex packings of fibers and hair. 

\begin{figure}
    \centering
    \includegraphics[width = 0.4\textwidth]{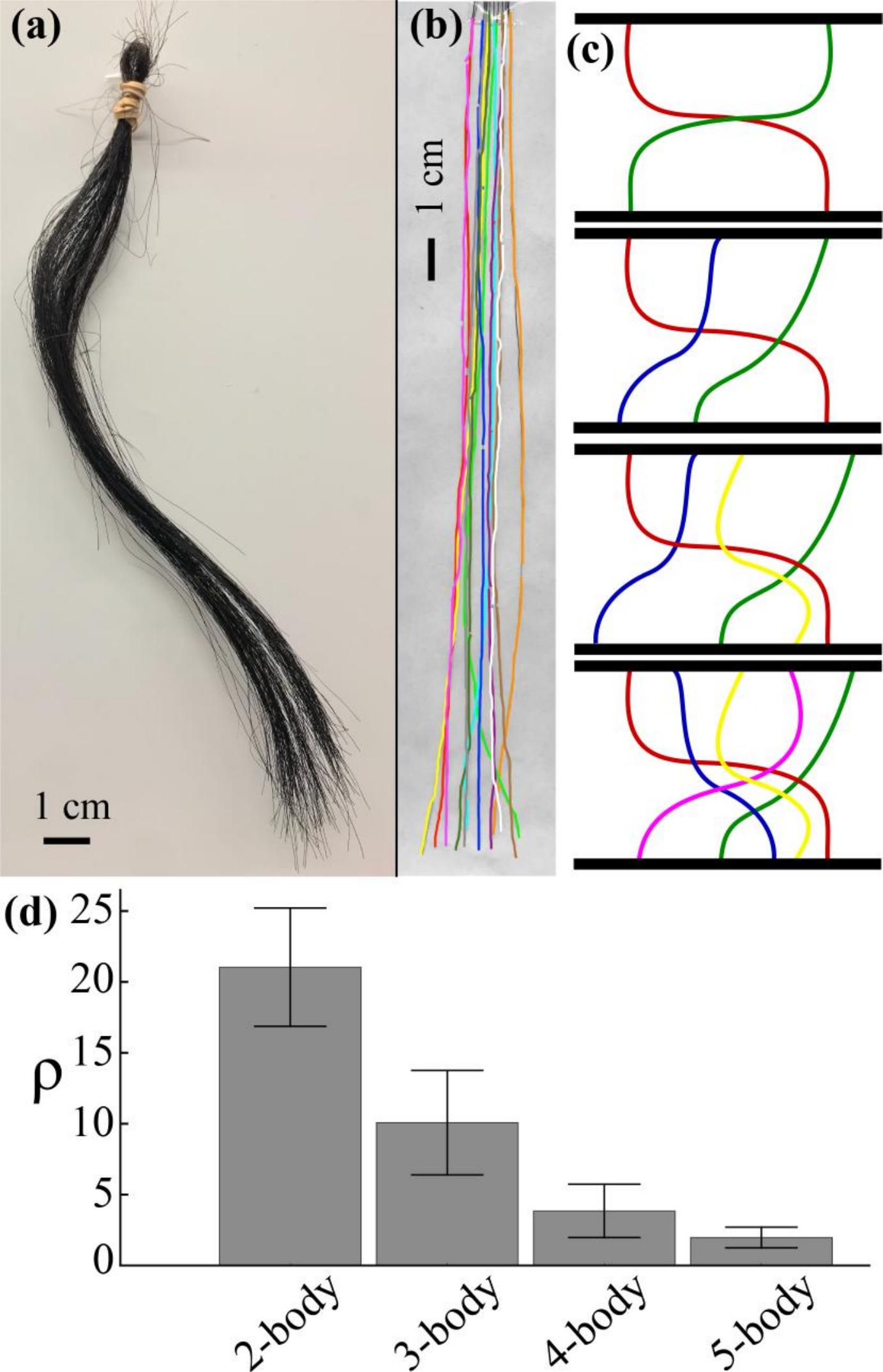}
    \caption{Tangles in hair. \textbf{(a)} Curl of horse hair. \textbf{(b)} Colored curl of 12 human hairs, clamped at one end. \textbf{(c)} Schematics of N-body interactions (\(N=2\)-\(5\)). \textbf{(d)} Histogram of N-body interactions for a sample similar to (b) with \( \rho = \) number of interactions per unit length (in meters). We segment the curl into 20 sections, count interaction types defined in \textbf{(c)} within each section and average.}
    \label{fig1}
\end{figure}

The complexity of the combing problem is potentially associated with the many-body nature of interacting filaments (hairs) and their potential for long-range interactions.  As an example, observations of a curl of horse hairs (Fig.~\ref{fig1}a) show that they are mostly straight, enhanced by gravity that helps straighten the hairs; only near the free end can the hairs adopt their natural curvature (see SI for detailed analysis).  To quantify the nature of these interactions, we digitally color each strand to track their interactions with their neighbors in a curl (Fig.~\ref{fig1}b),  segment the curl into sections, and then count and characterize the internal interactions of each strand using the interaction density in Fig.~\ref{fig1}c. We  define an interaction to be where strands cross each other, which is an upper bound on the number of interactions (since it does not account for the sign of the interaction, i.e. if a hair moves over or under another). In Fig.~\ref{fig1}d, we plot the number of N-body interactions, and see that pair-wise interactions form a plurality of the tangle types in our analyzed curls. 

\begin{figure}
\begin{center}
\includegraphics[width=0.45\textwidth]{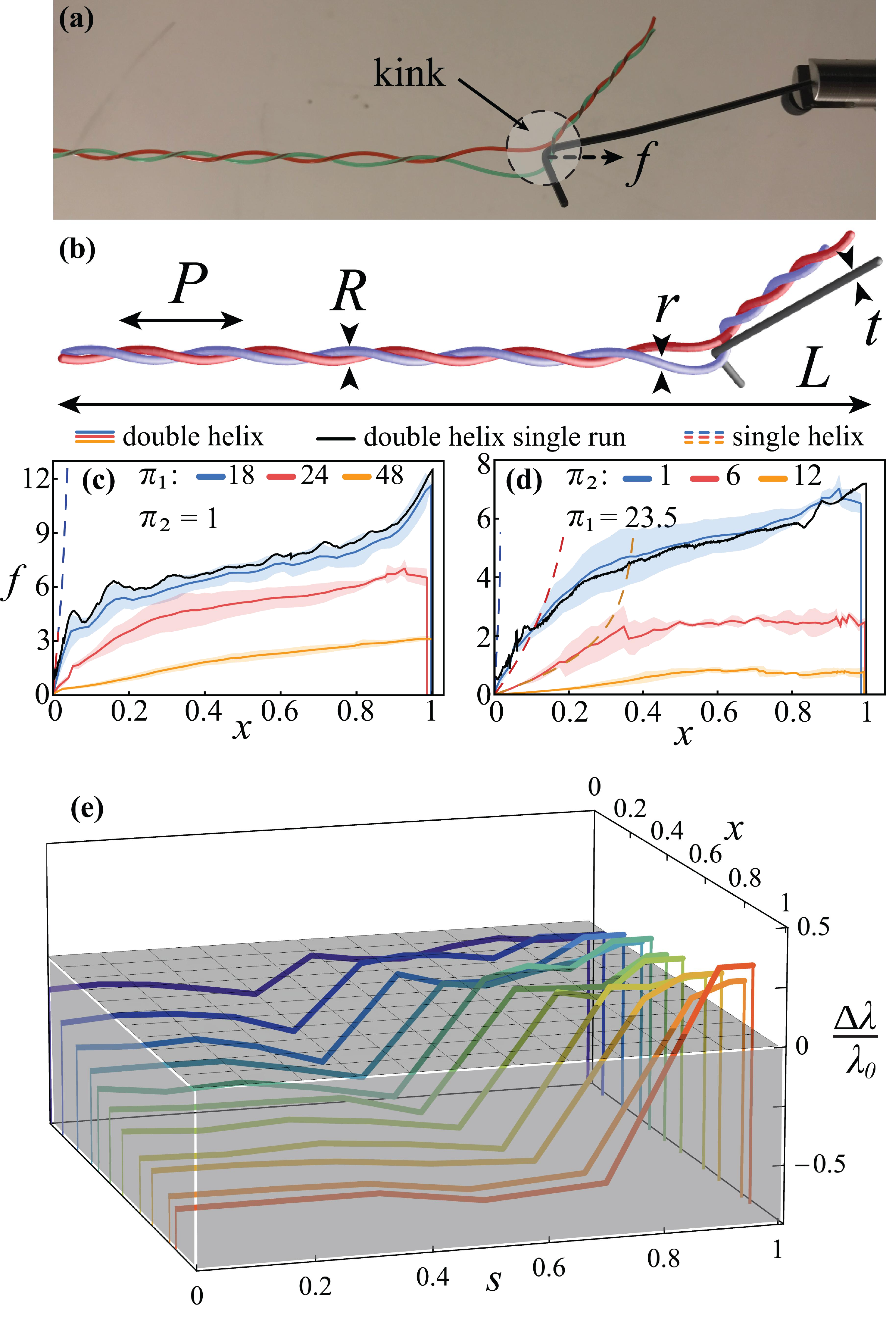}
\caption{Combing a double helix. \textbf{(a)} As the tine moves down the double helix, it  forms a kink about which the helix whirls as link flows out of the bundle.   \textbf{(b)} Schematic of combing a double helix with pitch \(P\), helix radius \(R\) (half the distance between individual filament centerlines), filament radius \(r\), tine radius \(t\) and helix height \(L\). \textbf{(c-d)} Scaled force $f = \frac{l_0^2 F}{B}$ applied by tine over scaled distance combed $x = \frac{d}{D}$, where $F$ is unscaled force, $d$ is unscaled distance combed, $l_0 = 0.01$m is a characteristic length, $B$ is single filament bending rigidity and \(D\) is total tine displacement required to detangle the helices. Dashed curves show theoretical force predicted for stretching a single helix; black curves show a single representative combing. We scan the phase space varying (\textbf{c}) \( \pi_1 = P/r\) and \textbf{(d)} \(\pi_2 = R/r\).  See SI for details and variation of \( \pi_3 = t/r\). \textbf{(e)} Change in link density $\Delta \lambda$ normalized by initial link density \(\lambda_0\) for \(\pi_1 = 18\), \( \pi_2=1\), plotted for several normalized tine locations, showing flow of link from clamped to free side and eventually out free end by twirling of the free ends about one another. In all experiments, \( r = 4.25 \cdot 10^{-4}\)~m and helix height \( h = 0.15\)~m. }
\label{fig2}
\end{center}
\end{figure}

Given the dominance of two-body interactions in a hair curl, we first consider a minimal model of the comb-curl system: two homochiral entwined helices clamped at the top end and hanging freely at the bottom. The filaments are made of nylon  (heated to force them to conform to 3-D printed helically grooved cylinders, and then cooled). This assembly is then pierced at the midpoint of the double helix centerline by a single stiff rod (the comb) which moves downward to detangle the curl.  We use an Instron 5566 material testing machine to measure the force extension curves of the rod-helix combing system. In Fig.~\ref{fig2}a we show the response of the helix that can be associated with either kinking or winding. Kinking arises when there is some sliding and shear between the free-end sides of the two helices, leading to a characteristic bent state (Fig~\ref{fig2}a,b). Over- and under-winding describes the stretching out of  the helices behind the tine and a concomittant compression of the free end.  As the tine moves through and along the double helix, the force-extension profile of the process evolves. We examine this in terms of the helix radius \( R \), filament radius \( r \), tine radius \( t \), and helix pitch \( P \) (Fig. \ref{fig2}b), expressed in terms of the dimensionless ratios \(\pi_1 = P/r\), \( \pi_2 = R/r \) and \( \pi_3 = t/r\), deferring variation of \( \pi_3 \) to the supporting information since \( t \) appeared to have little impact on the nature of helix unwinding.  

\begin{figure}
\begin{center}
\includegraphics[width = \columnwidth]{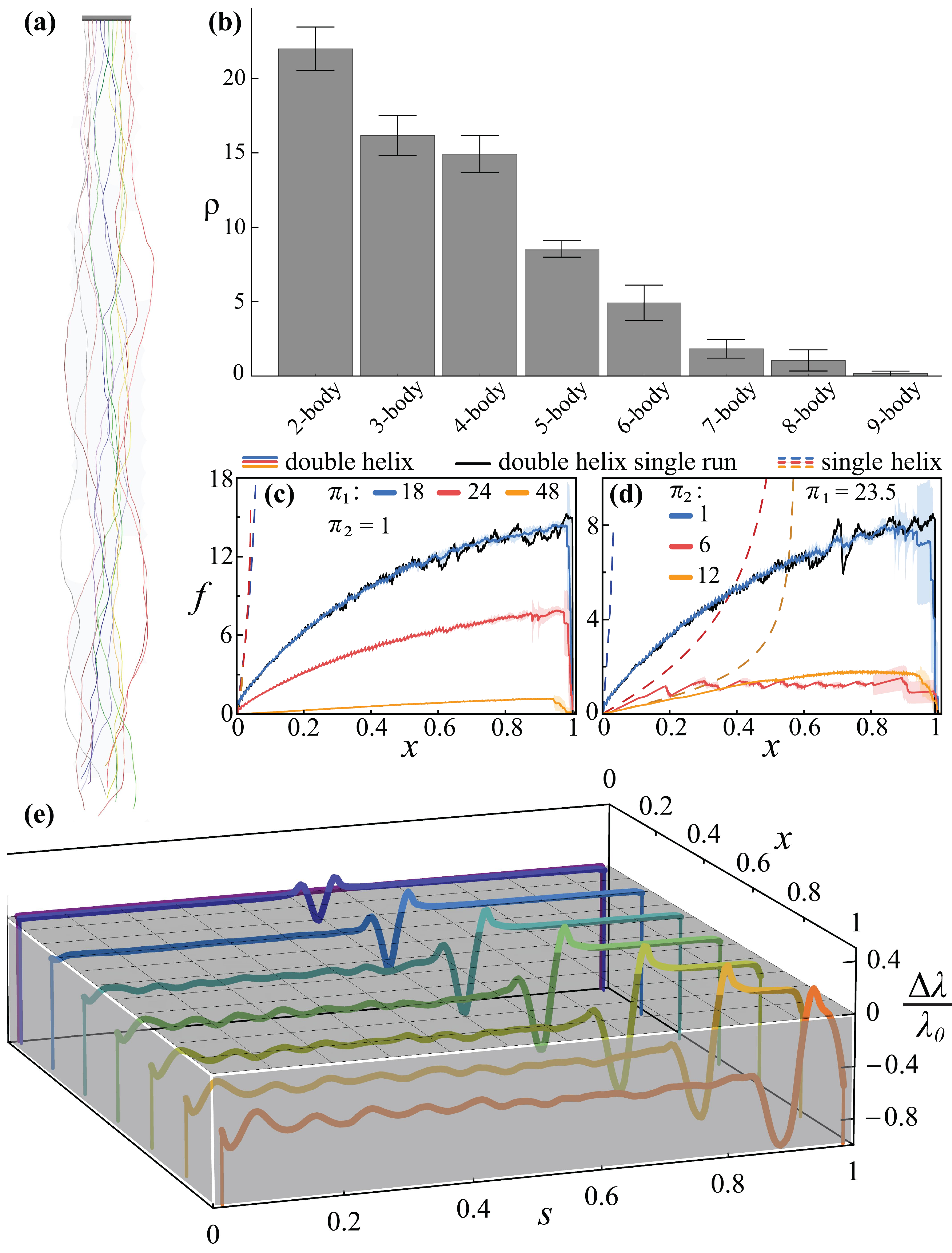}
\caption{Numerical results. \textbf{(a)} Simulated curl of 12 hairs modeling those in Fig.~\ref{fig1}b. \textbf{(b)} Hair interactions as in Fig. \ref{fig1}d for simulations similar to (a), where \(\rho = \) number of interactions per unit length (in meters). \( L_0 = 0.2\)~m, \( r = 7.5 \cdot 10^{-5}\)~m and \( E = 1\)~GPa. See SI for 3-dimensional curl results. \textbf{(c-d)} Scaled force $f = \frac{l_0^2 F}{B}$ applied by tine over scaled distance combed $x = \frac{d}{D}$, where parameters are defined as in Fig. \ref{fig2}. Dashed curves show theoretical force predicted for stretching single helix an equivalent distance; black curves show single representative combing. We scan the phase space varying (\textbf{c}) \( \pi_1 = P/r\) and \textbf{(d)} \(\pi_2 = R/r\).  See SI for details and variation of \( \pi_3 = t/r \). \textbf{(e)} Change in link density $\Delta \lambda$ normalized by initial link density \(\lambda_0\) for helix with \( \pi_1 = 18\), \( \pi_2 = 1 \), plotted for several tine locations, qualitatively reproducing experimental trends from Fig. \ref{fig2}e. Filament parameters are same as in Fig.~\ref{fig2}, except \(E\) values are scaled down by 0.078.  See SI for simulation settings.}
\label{fig3}
\end{center}
\vspace{-5mm}
\end{figure} 

In Fig.~\ref{fig2}d,e we show the scaled force on the tine \( f=F l_0^2/B \) as a function of the scaled tine displacement \( x \). Smaller pitch and radius helices (more tightly wound) lead to larger forces required to detangle. There is generally an initial rise in the force extension curve before a leveling off, corresponding to overwinding of the helix in front of the tine. After the initial rise, the force peaks when the tine jams, falling once the tine breaks through, although there is no rise in the force or tine jamming for loose helices, i.e. \( P/r > c \approx\)\(20\). We overlay the force required to equivalently stretch a single helix for comparison (initial angle \( \alpha \) and helix radius \( R \) extended to final radius angle \( \alpha_1 \) and helix radius \( R_1 \) \cite{Love}; see SI for details). In contrast with the force-extension curve for a single helix wherein the force required to stretch diverges as the helix is straightened out, combing through a double helix shows that the force flattens out before either helix is significantly stretched. Instead, the moving tine untangles the two filaments, eventually decoupling the helix strands as the force becomes vanishingly small. 

The unlinking of the homochiral helices during this process can be quantified in terms of the Calugareanu-Fuller-White (CFW) theorem \cite{Fuller1971, Calugareanu1959, White1969} which states that $Lk = Tw+Wr$, where Link (\(Lk\)) quantifies the oriented crossing number of the two filaments averaged over all projection directions (or Gauss Linking integral) and effectively counts the number of full turns one filament makes around the other \cite{Ricca2011}, Twist (\(Tw\)) is the integrated rotation of one filament around the two-filament-centerline, and Writhe (\(Wr\)) is the (negative) integral of the geometric torsion of the centerline.  To quantify the topology of the double helix, we treat the two interwoven filaments as two edges of a ribbon, and follow the local link density \( \lambda(s) \), defined as link per unit length along the double helix centerline as a function of centerline arclength \( s\), and the twist density \( \tau(s) \). For a relatively straight double helix, $Wr\approx 0$, so that  \( \lambda(s) \approx \tau(s)\) \cite{Kamien1997}. In Fig.~\ref{fig2}e we show the evolution of the link density \(\lambda\)  calculated from images of the helix taken as the tine moved. We see that a relatively uniform \( \lambda(s) \) changes to a step-like \( \lambda(s) \) as the tine induces a flow of link toward the free end of the curl. This leads to overwinding in front of the tine and underwinding behind while link flows through the helical braid from the clamped to the free end.  For tine displacements \( x \gtrsim 0.2 \),  the scaled jump in the link density \( \Delta \lambda/\lambda_0 > 0 \) across the tine (\(  \lambda_0 \) being the initial link density) increases faster than the rate it which is expelled from the free end.  This accumulation eventually reaches a plateau as the tine gets closer to the free end, and then induces a flux of link at the free end that unlinks the filaments, the end goal of combing.   

We now turn to correlate the experimentally observed spatio-temporal evolution of link density associated with the motion of the tine to the evolution of the force on the tine.  For a double helix that is tightly wound initially, as the tine propagates, \( \lambda(s) \) ahead of the tine increases, and the scaled pitch \( \pi_1 = P/r\) decreases locally. This causes the double helix to stiffen and eventually the tine jams when the local link density is larger than a threshold that depends on the diameter, the nominal helix pitch and radius of the filaments as well as the coefficient of friction. When one of the filaments slides relative to another, the relative shear often leads to the kinked configuration seen in Fig.~\ref{fig2}a,b accompanied by a peak in applied force. If the filament does not break, eventually the tine breaks through and the force decreases, before the same cycle repeats again. During this process, the kinked double helix twirls about the axis of tine motion (see SI video-S1). If the double helix is loose enough initially, link propagates more easily from the tine location to the free end, \( \lambda(s) \) never crosses the threshold needed for kinking, and the force required to comb never peaks (as seen for \( \pi_1 = 48, \pi_2 = 1\)). These are the only ways by which link flows out of the free end, via {untwisting} of the free ends of the individual filaments, and via the twirling rotation of the kinked portion of the double helix.  

 
To quantify our experiments on the non-linear topological mechanics of interacting filaments, we use  a numerical approach that models each hair using the Kirchhoff-Cosserat theory \cite{Cosserat1909,OReilly2017} and solve the equations using the discretization, contact-force and numerical integration scheme described in \cite{Gazzola2016}.   In the limit of very thin filaments such as hair, the model naturally reduces to Kirchhoff-Love theory for inextensible, unshearable filaments \cite{Love}.  We define \( s \in [0, L_0] \) as the material coordinate (also the arc length) of the rod of rest length \( L_0 \), $\B{{x}}(s)$ as the position vector of the center-line,  and a triad of orthonormal directors $\B{{d}}_1(s), \B{{d}}_2(s), \B{{d}}_3(s)=\partial_s \B{{x}}$ that defines the cross-section orientation. Then any body-convected vector \( \B{v} \) with lab-frame coordinates \( \B{\bar{v}} \) may be written as $\B{v} = \B{Q} \B{\bar{v}}$, where $\B{Q}(s) \in SO(3)$ is a rotation matrix, and the bending and twist strain vector is given by \( \B{\kappa} = {\rm vec} (\partial_s \B{Q}^T \B{Q}) \) (\(\B{\kappa}_0\) is rest-state curvature). If \( \B{N} \)(s) is the internal force resultant,  \( \B{f}_g \) as gravitational force line density, the equilibrium equations for the boundary value problem are \cite{OReilly2017, Gazzola2016}
\begin{align}
0 =& \partial_s \B{{N}} + \B{{f}}_g  \\
0 =& \partial_s \left( \B{B} \left( \B{\kappa} - \B{\kappa}_0 \right) \right) +    \partial_s \B{{x}}   \times \B{n} 
\end{align}
where $\B{B}$ is the matrix of bending and twisting stiffnesses, subject to the boundary conditions of the filaments being clamped at one end and free at the other.  

To simulate the initial state of a curl of hairs such as that depicted in Fig.~\ref{fig1}b, we start with a collection of clamped filaments hanging in a gravitational field $ |\B{{f}}_g|= \rho A g$ ($g \approx 9.8$ N/kg, $\rho$ is the filament mass density and $A$ is the filament cross-sectional area), resulting in a curl such as that shown in Fig~\ref{fig3}a. We then introduce a intrinsic curvature at each node, randomly drawn from a Gaussian distribution with mean and variance matching the distribution of curvatures (see SI for details), shown in Fig.~\ref{fig1}b.  Finally, we let the hairs relax elastically to their new rest configurations, determined by a competition between nonzero intrinsic curvature and gravitational straightening (see SI).
After the hairs relax, we count interactions using the same method as used in experiments, leading to the histogram of interactions per unit length in Fig. \ref{fig3}b. We find that pairwise interactions dominate in agreement with experimental results (see SI where we show that these interactions do not change qualitatively in a 3d array). 

To follow the combing of an elastic double helix as in Fig.~\ref{fig2}, we start with a random distribution of initial internal strains and anneal the pair of filaments into a double helix, clamped at one end. We then insert a rigid rod as a tine between the two helices close to the clamped end and move it quasi-statically towards the free end. In Fig.~\ref{fig3}c-d we show the numerically computed force applied by the tine during combing, and see that the results are similar to our experiments shown in Fig.~\ref{fig2}c-d \footnote{The high-frequency oscillations seen in the mean force curves are due to the oscillation in contact force between the tine and double helix filaments as the tine passes by each discretization node in the double helix filaments. }. Comparing the evolution of the helix topology computed using the algorithms described in \cite{solenoids}, Fig.~\ref{fig3}e shows that this also matches the experimental pattern of overwinding and underwinding shown in Fig.~\ref{fig2}e; the oscillations in \( \lambda(s)\) on the clamped side of the tine come from slight periodic shearing of the clamped side filament segments with respect to each other, a phenomenon that occurs on too fine a resolution to be seen in experiments. The small discrepancies between the quantitative values observed in the numerical and experimental results may come from our scaling of the simulated filaments' bending stiffness \footnote{Note that the numerical calculations presented here neglect contact friction.  However, we tested a subset of these simulations with contact friction and obtained force-displacement curves with the same shape as the ones shown here.  Hence, for simplicity, we neglect contact friction in all simulations shown here.}.  
 
\begin{figure*}
\begin{center}
    \includegraphics[width=\textwidth]{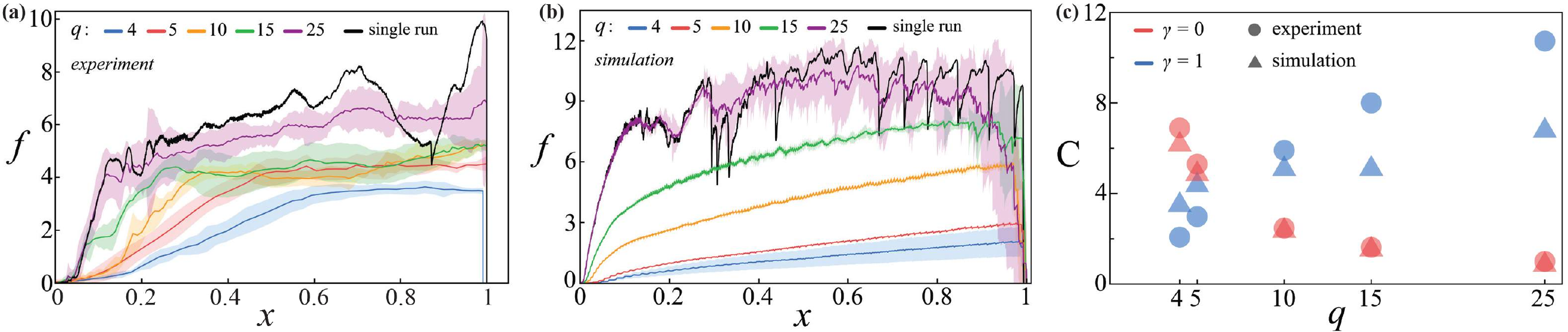}
    \caption{\textbf{(a-b)} Scaled force \(f\) over scaled distance combed $x$ for \textbf{(a)} experimental and \textbf{(b)} simulated double helices with varying number of links. Scaling of \( x \) is different for each curve since the total raw tine displacement needed to comb each curl is different. Below a certain link threshold, combing finishes without kink formation.  Black curves show a single representative combing for \(q = 25 \).  Filament parameters are the same as in Fig.~\ref{fig2}.  \textbf{(c)} Cost \(C\) to comb a 10m double helix over number of downstream pitches \( q \) detangled per combing, using the strategies shown in (a-b). Note that \( q \) is intrinsic to combing strategy.  \(C\) combines effective pain and time required to comb; \( \gamma = 0 \) considers only time and corresponds to straight hair; \( \gamma = 1\) considers only pain and corresponds to curly hair. Simulation and experiment how optimal combing strategy shifts depending on hair curvature.}
    \label{fig4}
\end{center}
\vspace{-5mm}
\end{figure*} 

Complementing our analysis of the topological mechanics of combing a double helix, we now consider how a curl's resistance to combing varies with the style of combing, e.g. short strokes versus long strokes. Noting that a single helix pitch corresponds to one helix filament revolving once around the other helix filament, we fix \( R/r = 1 \) and \( P/r = 24\) and vary \( q \), the number of pitches that initially separate the tine from the free end of the double helix.  In experiment (Fig.~\ref{fig4}a) and simulation (Fig.~\ref{fig4}b) we find that significant tine jamming---as indicated by the plateaus in the force-displacement curves---occurs only when we start combing more than 4-5 helix pitches away from the free end \footnote{Quantitative differences in the shape of experimental and numerical force-displacement curves in Fig.\ref{fig4} likely stem from our model's neglect of contact friction.}.The differences between the two sets of curves is likely due to not accounting for friction in the computations, and explains the larger number of plateaus seen in the large-\(q\) experiments relative to the more smooth force evolution seen in computations. Nevertheless, we see that starting to comb nearer to the free end allows link to be expelled more easily, leading to complete untangling before the differential link density across the tine surpasses the threshold for kink formation and jamming, consistent with experience. 

To quantify this intuitive result, we define a cost $C(q)$ for combing a double helix of length $L$ using each strategy, balancing the effects of  \( f_{\text{max}}\), the maximum dimensionless force during combing, and the amount of detanglement quantized in terms of the number of strokes $q_0/q$ need to complete the combing process, where $q_0=25$ a fixed number of pitches characteristic to the \( q \)-values used in these combing strategies. Writing $C(q)= \gamma f_{\text{max}} + ( 1 - \gamma ) \frac{q_0}{q}  $ ( see SI for details), we can vary the relative cost of the two effects by varying $\gamma$. Varying \( \gamma \) allows us to optimize a combing strategy based on the initial nature of the double helix; for very curly hair, choosing $\gamma \sim O(1)$  is more sensible, while for straight hair, choosing $\gamma \sim 0$ is better. Using the results for the maximum force from Fig.~\ref{fig2}-\ref{fig3}, we calculate the cost $C(q)$ for the case of curly and straight hair. In Fig. \ref{fig4}c, we see that for straight hair ($\gamma \sim 0$), the cost decreases as $q$ increases, i.e. the number of strokes decreases, while for curly hair ($\gamma \sim 1$), the cost is lowest for small $q$ as it is biased to minimize the cost associated with the maximum force.

 
Our study has shown that the many-body problem of combing hair can be simplified to the problem of combing a double helix, and in this context, we have provided a  measure of link density during combing as a function of helix geometry and combing procedure.  We have connected topology, geometry and mechanics by quantifying the relation between flow of link  through the tine and out the free end, to the time-varying force felt on the tine.  Our results also suggest that the two-body problem also has the ability to capture the correct optimal strategy of combing a tangle by balancing the cost of many short strokes relative to longer, potentially more painful ones. Going forward, a natural problem is to account for the strong frictional anisotropy of the hairs as well as their response to combing as a function of humidity and temperature. 

 {Acknowledgements. For partial financial support, we thank the National Science Foundation grants NSF DMR 20-11754, NSF DMREF 19-22321, and NSF EFRI 18-30901.}


\begin{thebibliography}{1}

\bibitem{Hearle}
W. E. Morton and J. W. S. Hearle, \textit{Physical Properties of Textile Fibres}, (Textile Institute, Manchester, 1993). 


\bibitem{Pan}
N. Pan, \textit{Text. Res. J.} \textbf{62} (12), 749 (1992).

\bibitem{Panaitescu}
A. Panaitescu, G. M. Grason, and A. Kudrolli, \textit{Phys. Rev. E} \textbf{95} (5-1), 052503 (2017). 

\bibitem{Kabla}
A. Kabla and L. Mahadevan, \textit{J. R. Soc. Interface} \textbf{4} (12), 99 (2007). 

\bibitem{Leaf}
G. A. V. Leaf and W. Oxenham, \textit{Journ. of the Text. Inst.} \textbf{72} (4), 168 (1981). 


\bibitem{Goldstein2018}
P. Warren, R. Ball, R. Goldstein, \textit{Phys. Rev. Lett.} \textbf{120} (15), (2018). 

\bibitem{Durand2009}
G. Gurtner and M. Durand, \textit{EPL} \textbf{87}, 24001 (2009). 

\bibitem{Sacks2017}
J. Carleton, G. Rodin and M. Sacks, \textit{Acs Biomaterials Sci. \& Eng.} \textbf{3} (11), 2907-2921 (2017). 

\bibitem{Ward2007}
K. Ward, F. Bertails, Tae-Yong Kim, S. R. Marschner, M.-P. Cani and M. C. Lin. \textit{IEEE Trans. Visualiz. \& Computer Graphics} \textbf{13} (2), 213-234 (2007).

\bibitem{Bao2017}
Yongtang Bao and Yue Qi. \textit{IEEE Access} \textbf{5}, 12533-12544 (2017). 

\bibitem{Love}
A. E. H. Love, \textit{A Treatise on the Mathematical Theory of Elasticity}, 2nd ed., vol. 2, (University Press, Cambridge, 1906).

\bibitem{Ricca2011}
R. Ricca and B. Nipoti, \textit{Journ. Knot Theory and its Ramifications} \textbf{20} (10), 1325 (2011).

\bibitem{Fuller1971}
F. B. Fuller, \emph{Proc. Natl. Acad. Sci. USA} \textbf{68} (4), 815 (1971).

\bibitem{White1969}
James H. White, \textit{Am. J. Math} \textbf{91} (3), 693 (1969). 

\bibitem{Calugareanu1959}
G. Calugareanu, \emph{Rev. Math. Pures Appl.} \textbf{4}, 5 (1959).

\bibitem{Kamien1997}
R. D. Kamien, \textit{The Europ. Phys. Journ. B - Cond. Mat. \& Complex Systems} \textbf{1} (1), 1 (1998). 

\bibitem{Goldstein2000}
C. W. Wolgemuth, T. R. Powers and R. E. Goldstein, \textit{Phys. Rev. Lett.} \textbf{84} 1623 (2000). 

\bibitem{Cosserat1909}
E. Cosserat and F. Cosserat. Theorie des corps deformables. Paris. (1909). 

\bibitem{OReilly2017}
O. M. O'Reilly, \textit{Modeling nonlinear problems in the mechanics of strings and rods: the role of the balance laws} (Springer,  2017).

\bibitem{Gazzola2016}
M. Gazzola, L. Dudte, A. McCormick and L. Mahadevan, \textit{R. Soc. Open Sci.} \textbf{5}, 171628 (2018). 

\bibitem{solenoids}
N. Charles, M. Gazzola and L. Mahadevan, \textit{Phys. Rev. Lett.} \textbf{123}, 208003 (2019).  

\end{thebibliography}
\end{document}